\documentclass[12pt]{article}
\usepackage{amsfonts}
\usepackage{amsmath}
\usepackage{amssymb}
\usepackage{graphicx}
\usepackage{color}
\usepackage[left=1cm,top=1cm,bottom=1cm,right=1cm,nohead,nofoot]{geometry}

\def \be {\begin{equation}}
\def \ee {\end{equation}}
\def \bea {\begin{eqnarray}}
\def \eea {\end{eqnarray}}
\def \nn {\nonumber}

\def \rr {\raise.35ex\hbox{\small $\prime$}\kern-.17em{\mbox{\large $\imath$}}}
\def \del {\partial}
\def \dels {\partial\kern-.6em /\kern.1em}
\def \As {{A\kern-.5em / \kern.5em}}
\def \Ds {D\kern-.7em / \kern.5em}
\def \a {\alpha}

\def \b {\beta}

\def \g {\gamma}
\def \G {\Gamma}
\def \d {\delta}
\def \eps {\epsilon}
\def \m {\mu}
\def \n {\nu}

\def \ks {k\kern-.5em /}
\def \ls {l\kern-.5em /}\def \lam {\lambda}

\def \II {I\hspace{-.1em}I\hspace{.1em}}

\def \IIA {\mbox{\II A\hspace{.2em}}}

\def \dd {\dot{\delta}}

\def \dm {\dot{\mu}}
\def \dn {\dot{\nu}}

\def \dlam {\dot{\lambda}}
\def \ds {\dot{\sigma}}
\def \dr {\dot{\rho}}
\def \dt {\dot{\tau}}

\reversemarginpar
\newcommand{\ol}{\overline}

\renewcommand{\thefootnote}{\fnsymbol{footnote}}

\begin{document}
\begin{titlepage}

\begin{center}

\hfill
\vskip .2in

\textbf{\LARGE
 Supersymmetry and BPS States on\vskip.5cm
   D4-brane in Large $C$-field Background
}

\vskip .5in
{\large
Chen-Te Ma$^{a,}$\footnote{e-mail address: yefgst@gmail.com} and\
Chi-Hsien Yeh$^{a,b, }$\footnote{e-mail address: d95222008@ntu.edu.tw}}\\
\vskip 6mm
{\sl
${}^a$
Department of Physics and Center for Theoretical Sciences,\\
National Taiwan University, Taipei 10617, Taiwan, R.O.C.\\
\vskip 3mm
${}^b$ 
National Center for Theoretical Sciences, \\
National Tsing-Hua University, Hsinchu 30013, Taiwan,
R.O.C.}\\
\vskip 3mm
\vspace{60pt}
\end{center}
\begin{abstract}
In this paper, we continue our previous study of the low energy effective theory for D4-brane in the large $C$-field background. The gauge field part of the effective action was found in an earlier work.  In this paper, we focus on the matter field part of the action and the supersymmetry transformation. Moreover, we calculate the central charges of super algebra and extensively study BPS solutions of this effective theory. 
The BPS states considered in this paper include
light-like gauge field configurations,
the F1 ending on D4 solution,
tilted D4-brane, BPS solution with two types of magnetic charges (D2 ending on D4), 
holomorphic embedding of D4-brane
and the intersection of two D4-branes along a 2-brane.

\end{abstract}

\end{titlepage}

\renewcommand{\thefootnote}{\arabic{footnote}}
\setcounter{footnote}{0}

\section{Introduction}
\label{1}
The descriptions of M-theory are five superstring theories and 11-dimensional supergravity. They are related to each other from duality and dimension reduction  \cite{Witten:1995ex}. For example, the \IIA superstring theory can be corresponded to the 11-dimensional supergravity compactified on $S^{1}$ circle \cite{Townsend:1995kk}. The solitonic solutions of 11-dimensional supergravity are called M2 and M5 branes, which play important roles as D-branes in string theory. Although the description of M-theory is not known totally, the low energy effective theory of two and five branes are known to be.  

A recent progress was
the construction of the low energy effective theory 
for a single M5-brane in the large $C$-field background \cite{M51,M52,Ho:2009zt} \footnote{The large $C$-field limit for M5-brane was first considered in \cite{Berman}.} and its 1/2 BPS states \cite{Ho:2012dn}.
We call this theory to be the Nambu-Poisson M5-brane theory or NP M5 theory in short and all known 1/2 BPS states have their counterparts in the absence of $C$-field background.
From the NP M5-brane theory, we can derive the low energy effective theory for D4-brane in the large $C$-field background from double dimensional reduction (DDR) method \cite{Ho}.

This D4-brane effective action for a large $C$-field background is different from the original DBI description \cite{Leigh}. This new D4-brane action has Nambu-Poisson structure as the NP M5-brane theory. As it is well known, the effective action of D-brane in the large NS-NS $B$-field background should be described by gauge theories on noncommutative space \cite{ChuHo,Schomerus,Seiberg:1999vs}. A similar phenomenon also occurs in the low energy effective theory for D4-brane in the large $C$-field background.  The gauge symmetry
of an NP M5-brane is the volume-preserving diffeomorphism (VPD) \cite{Matsuo}, which is described by the Nambu-Poisson bracket.  After doing DDR on the worldvolume direction which is different with the $C$-field
background directions, we get an effective D4-brane with the $C$-field. Since the $C$-field
background is parallel to the D4-brane after DDR, it is natural to expect that the D4-brane inherits the VPD symmetry. It is also expected that the geometry of this theory is equipped with a 3-bracket structure \cite{Chu:2009iv,Huddleston:2010cx}. 

One purpose of this paper is to study more details of this effective action with matter fields. The supersymmetry transformation of all fields and the modification of field strengths from matter fields will be given in this paper. In the previous paper \cite{Ho}, we did not write down the manifest supersymmetry transformation of all fields. The reason is that we did not only do DDR from the NP M5-brane theory but also do electric-magnetic dual transformation of this theory. The supersymmetry transformation of new fields in dual action cannot be directly obtained from DDR on the NP M5-brane. Hence, we need to completely calculate the supersymmetry transformation of this dual action, and find the supersymmetry transformation of these new fields to make the dual action invariant. In this paper, we give all supersymmetry transformation of fields for this effective theory, and also hope to give a starting point on the construction of D$p$-brane in the R-R $(p-1)$-form field background. 

 In this paper, we also want to find 1/2 BPS states
on D4-brane in the large $C$-field background. Most of solutions on D4-brane are easy to be expected from M5-brane. To simplify, we do not write down all details for these solutions. The BPS configurations considered in this paper include
light-like solutions, the F1 string ending on D4-brane,
tilted D4-brane, BPS solution with two types of magnetic charges, 
holomorphic embedding of D4-brane
and the intersection of two D4-branes along a 2-brane.

The plan of this paper is as follows.
We review the D4-brane in the large $C$-field background in Sec. \ref{2} and the gauge symmetry and supersymmetry in Sec. \ref{3}.
In Sec. \ref{4} we show BPS configurations on D4-bane. But we do not show the details on the solutions derived from the NP M5-brane theory for brevity.  
Finally, in Sec. \ref{5} we conclude.

\section{D4-Brane in Large $C$-field Background}
\label{2}
To carry out DDR
on the NP M5-brane theory along the $x^2$-direction \footnote{The NP M5-brane theory is defined on six dimensional worldvolume, and we use the notation  $\{x^{0},x^{1},x^{2};x^{\dot{1}},x^{\dot{2}},x^{\dot{3}}\}\equiv\{x^{\mu};x^{\dm}\}$ to label these worldvolume coordinates. The $C$-field background is given by $C_{\dot 1\dot 2\dot 3}dx^{\dot1}dx^{\dot2}dx^{\dot3}$.},
we set
\be
x^2 \sim x^2 + 2\pi R,
\ee
and let all other fields to be independent of $x^2$.
As a result we can set $\del_2$ to be zero when it acts on any fields.
Here $R$ is the radius of the circle of compactification
and we should take $R \ll 1$
to let M5-brane
reduce to D4-brane.

After we perform DDR on the $x^2$ direction, we need to re-interpret the gauge fields $b_{\mu\dm}$ and $b_{\dm\dn}$ in the NP M5-brane theory from the viewpoint of D4-brane. 
The gauge field $b_{\m\dm}$ becomes two kinds of fields $\{b_{2\dm},b_{\a\dm}\}$ after DDR, where $\a=0,1$. We should identity $b_{2\dm}$ as components of the one-form potential on the D4-brane,
\be
b_{\dm2}\equiv a_{\dm}.
\ee
We use the same chirality condition with the NP M5-brane theory,
\be
\G^{012\dot{1}\dot{2}\dot{3}}\Psi=\Psi ~~,~ \G^{012\dot{1}\dot{2}\dot{3}}\eps=-\eps.
\ee

\subsection{Action}
After performing DDR on the $x^2$ direction, we get the low energy effective description for D4-brane in the large $C$-field background \cite{Ho}.
The action of the gauge fields is
\bea
S^{D4}_{gauge}&=&\int d^{5}x
\left\{
-\frac{1}{2}{\cal H}_{\dot{1}\dot{2}\dot{3}}^{2}
-\frac{1}{4}{\cal H}_{2\dm\dn}^2
\right. \nn\\
&&-\frac{1}{4}{\cal H}_{\a\dm\dn}^2
\left. +\eps^{\a\b}\del_{\b}a_{\dm}B_{\a}^{~\dm}
+\frac{g}{2}\eps^{\a\b}F_{\dm\dn}B_{\a}^{~\dm}B_{\b}^{~\dn}\right\},\label{D4inCgauge}
\eea
where we use the definition, $\eps^{\a\b2}\equiv\eps^{\a\b}$.

The action of the scalar fields $X^{I} ~(I=6,7,8,9,10)$ is
\bea
S^{D4}_{X}&=&\int d^{5}x
\left\{
-\frac{1}{2}{\cal D}_{\dm}X^{I}{\cal D}^{\dm}X^{I} -\frac{1}{2}\del_{\a}X^{I}\del^{\a}X^{I}+gB_{\a}^{~\dm}\del_{\dm}X^{I}\del^{\a}X^{I} \right.
\nn\\
&&- \frac{g^{2}}{2}B_{\a}^{~\dm}B^{\a}_{~\dn}
\del_{\dm}X^{I}\del^{\dn}X^{I}
-\frac{g^{2}}{8}\eps^{\dm\dr\dt}\eps_{\dn\ds\dd}F_{\dr\dt}F^{\ds\dd}
\del_{\dm}X^{I}\del^{\dn}X^{I}
\nn\\
&&\left.-\frac{1}{2g^2}-\frac{g^{4}}{4}\{X^{\dm},X^{I},X^{J}\}^{2}
-\frac{g^{4}}{12}\{X^{I},X^{J},X^{K}\}^{2}\right\}.\label{D4inCX}
\eea
The action of the fermionic field $\psi$ is
\bea
S^{D4}_{\Psi}&=&\int d^{5}x
\left\{\frac{i}{2}\bar{\Psi}\Gamma^{\a}\del_{\a}\Psi
+\frac{i}{2}\bar{\Psi}\Gamma^{\dr}{\cal D}_{\dr}\Psi
+g\frac{i}{4}\bar{\Psi}\Gamma^{2}\eps^{\dm\dn\dr}F_{\dn\dr}\del_{\dm}\Psi
-g\frac{i}{2}\bar{\Psi}\Gamma^{\a}B_{\a}^{~\dm}\del_{\dm}\Psi\right.
\nn\\
&&\left.+g^2\frac{i}{2}\bar{\Psi}\Gamma_{\dm}\G^{I}\{X^{\dm},X^{I},\Psi\}
-g^2\frac{i}{4}\bar{\Psi}\Gamma^{IJ}
\Gamma_{\dot{1}\dot{2}\dot{3}}\{X^{I},X^{J},\Psi\}\right\}.\label{D4inCPsi}
\eea

The Nambu-Poisson bracket $\{\cdot, \cdot, \cdot\}$
is used to define the algebraic structure for gauge symmetry.
In general, it satisfies Leibniz rule and fundamental identity.
It is defined by
\be
\{f, g, h\} = \eps^{\dm\dn\dlam}\del_{\dm}f \del_{\dn}g \del_{\dlam}h.
\ee

In the above, we used the notation
\bea
B_{\a}{}^{\dm} &\equiv& \eps^{\dm\dn\dr}\del_{\dn}b_{\a\dr},\\
b^{\dm} &\equiv& \frac{1}{2} \eps^{\dm\dn\dlam} b_{\dn\dlam}, \\
X^{\dot\mu} &\equiv&
\frac{x^{\dot\mu}}{g} + b^{\dm}.
\eea

The covariant derivatives are defined by
\bea
{\cal D}_\a\Phi
&\equiv&(\del_{\a} - gB_{\a}{}^{\dm}\del_{\dm}) \Phi,
\qquad
(\Phi = X^I, \Psi)
\label{dmu}
\\
{\cal D}_{\dot\mu}\Phi
&\equiv&\frac{g^2}{2}\epsilon_{\dot\mu\dot\nu\dot\rho}
\{X^{\dot\nu},X^{\dot\rho},\Phi\},
\eea
and the field strengths are defined by
\begin{eqnarray}
{\cal H}_{\a\dm\dn} &\equiv&
\eps_{\dm\dn\dlam}(\del_{\a}b^{\dlam}-B_{\a}{}^{\dlam}-g\del_{\ds}{b}^{\dlam}B_{\a}{}^{\ds}),\label{h12def}\\
{\cal H}_{\dot1\dot2\dot3}
&\equiv&g^2\{X^{\dot1},X^{\dot2},X^{\dot3}\}-\frac{1}{g}
\nonumber\\
&=&H_{\dot1\dot2\dot3}
+\frac{g}{2}
(\partial_{\dot\mu}b^{\dot\mu}\partial_{\dot\nu}b^{\dot\nu}
-\partial_{\dot\mu}b^{\dot\nu}\partial_{\dot\nu}b^{\dot\mu})
+g^2\{b^{\dot1},b^{\dot2},b^{\dot3}\}.
\label{h30def}
\end{eqnarray}

We also use the notation of Abelian field strength,
\be
F_{\dm\dn}=\del_{\dm}a_{\dn}-\del_{\dn}a_{\dm}.
\ee
In fact, the field strength of D4-brane is deformed by the $C$-field background. The field strength ${\cal H}_{2\dm\dn}$ can be understood as the deformed field strength on D4-brane,
\be
{\cal H}_{2\dm\dn}\equiv {\cal F}_{\dm\dn}.
\ee
The remaining two field strengths ${\cal F}_{\a\dm}$ and ${\cal F}_{\a\b}$ are also deformed by the $C$-field background. Before discussing this issue, we should study how to find the remaining one-form degree $a_{\a}$ in this theory.

\subsection{Electric-Magnetic Duality Transformation}
From the viewpoint of D4-brane, there are one-form gauge potential $a_{\dm}$ and $a_{\a}$.  The one-form gauge field $a_{\a}$ can be understood as the EM duality of $b_{\a\dm}$. After 3-dimensional EM dual transformation in $x^{\dm}$ spaces, the d.o.f of gauge field $b_{\a\dm}$ (one form in $x^{\dm}$ space) can be re-interpreted into the d.o.f of gauge field $a_{\a}$ (0-form in $x^{\dm}$ space). Moreover, the field $B_{\a}^{~\dm}$ can be understood as a new independent field $\breve{B}_{\a}^{~\dm}$ which is not divergenceless. The $\breve{B}_{\a}^{~\dm}$ is at most quadratic so it can be integrated out and re-expressed by the other fields ($b^{\dm}, a_{\dm}, a_{\a}, X^{I}, \Psi$) from the equations of motion.

The complete dual action is,
\begin{eqnarray}
\label{eq:2.1}
S[b^{\dm},a_A,\breve{B}_{\a}^{~\dm},X^{I},\Psi]&=&
\int d^{5}x\left\{
-\frac{1}{2}{\cal D}_{\dm}X^{I}{\cal D}^{\dm}X^{I} -\frac{1}{2}\del_{\a}X^{I}\del^{\a}X^{I}+g\breve{B}_{\a}^{~\dm}\del_{\dm}X^{I}\del^{\a}X^{I} \right.
\nn\\
&&- \frac{g^{2}}{2}\breve{B}_{\a}^{~\dm}\breve{B}^{\a}_{~\dn}
\del_{\dm}X^{I}\del^{\dn}X^{I}
-\frac{g^{2}}{8}\eps^{\dm\dr\dt}\eps_{\dn\ds\dd}F_{\dr\dt}F^{\ds\dd}
\del_{\dm}X^{I}\del^{\dn}X^{I}
\nn\\
&&-\frac{g^{4}}{4}\{X^{\dm},X^{I},X^{J}\}^{2}
-\frac{g^{4}}{12}\{X^{I},X^{J},X^{K}\}^{2}
\nn \\
&&+\frac{i}{2}\bar{\Psi}\Gamma^{\a}\del_{\a}\Psi
+\frac{i}{2}\bar{\Psi}\Gamma^{\dr}{\cal D}_{\dr}\Psi
+g\frac{i}{4}\bar{\Psi}\Gamma^{2}\eps^{\dm\dn\dr}F_{\dn\dr}\del_{\dm}\Psi
-g\frac{i}{2}\bar{\Psi}\Gamma^{\a}\breve{B}_{\a}^{~\dm}\del_{\dm}\Psi
\nn\\
&&+g^2\frac{i}{2}\bar{\Psi}\Gamma_{\dm} \G^I \{X^{\dm},X^{I},\Psi\}
-g^2\frac{i}{4}\bar{\Psi}\Gamma^{IJ}
\Gamma_{\dot{1}\dot{2}\dot{3}}\{X^{I},X^{J},\Psi\}
\nn\\
&&-\frac{1}{2g^{2}} -\frac{1}{2}({\cal H}_{\dot{1}\dot{2}\dot{3}})^{2}
-\frac{1}{4}{\cal F}_{\dn\dr}{\cal F}^{\dn\dr}
-\frac{1}{4}(\eps_{\dm\dn\dr}
(\del_{\a}b^{\dm}-V_{\ds}^{~\dm}\breve{B}_{\a}^{~\ds}))^2
\nn \\
&&\left.
+\eps^{\a\b}\del_{\b}a_{\dm}\breve{B}_{\a}^{~\dm}
+\frac{g}{2}\eps^{\a\b}F_{\dm\dn}\breve{B}_{\a}^{~\dm}\breve{B}_{\b}^{~\dn}-\eps^{\a\b}\del_{\dm}a_{\b}\breve{B}_{\a}^{~\dm}
\right\}.
\end{eqnarray}
Here, we define $V_{\dm}{}^{\ds}\equiv g \del_{\dm}X^{\ds}$.

$\breve{B}_{\a}^{~\dm}$ in the dual action is replaced by its EOM solution. The equation of motion for $\breve{B}_{\a}^{~\dm}$ is
\be
V_{\dm}^{~\dn}(\del^{\a}b_{\dn}-V^{\dr}_{~\dn}\breve{B}^{\a}_{~\dr})
+\eps^{\a\b}F_{\b\dm}+g\eps^{\a\b}F_{\dm\dn}\breve{B}_{\b}^{~\dn}
+g\del_{\dm}X^{I}\del^{\a}X^{I}-g\frac{i}{2}\bar{\Psi}\Gamma^{\a}\del_{\dm}\Psi
-g^{2}\breve{B}^{\a}_{~\dn}\del_{\dm}X^{I}\del^{\dn}X^{I}=0.
\label{r2}
\ee
Its solution is
\bea
\hat{B}_{\a}^{~\dm}&=& (\textbf{M}^{-1})^{\dm\dn}_{~~~\a\b}
(V_{\dn}^{~\ds}\del^{\b}b_{\ds}+\eps^{\b\g}F_{\g\dn}
+g\del_{\dn}X^{I}\del^{\b}X^{I}-g\frac{i}{2}\bar{\Psi}\Gamma^{\b}\del_{\dn}\Psi)
\nn \\
&\equiv&(\textbf{M}^{-1})^{\dm\dn}_{~~~\a\b}W_{\dn}^{\b},\label{hatB}
\eea
where
\be
\textbf{M}_{\dm\dn}^{~~\a\b} \equiv
(V_{\dm\dr}V_{\dn}^{~\dr}+g^{2}\del_{\dm}X^{i}\del_{\dn}X^{i})\d^{\a\b}
-g\eps^{\a\b}F_{\dm\dn}
\ee
and $(\textbf{M}^{-1})^{\dm\dn}_{~~~\a\b}$ is defined by
\be
(\textbf{M}^{-1})^{\dlam\dm}_{~~~\g\a}\textbf{M}_{\dm\dn}^{~~\a\b}
=\d^{\dlam}_{~~\dn}\d^{~\b}_{\g}.
\ee

In this dual action, the field $\breve{B}_{\a}{}^{\dm}$ is not divergenceless. We expect that there are remaining terms proportional to $\del_{\dm}\breve{B}_{\a}{}^{\dm}$ after we do gauge and supersymmetry transformation except for $a_{\a}$ and $\breve{B}_{\a}{}^{\dm}$. (We use dimensional reduction to obtain the transformation laws of other fields from M5-brane to D4-brane.) To make the dual action being invariant, we should find gauge and supersymmetry transformation of $a_{\a}$ and $\breve{B}_{\a}{}^{\dm}$ to cancel these remaining terms. In next section, we show the details of gauge and supersymmetry transformation of $a_{\a}$ and $\breve{B}_{\a}{}^{\dm}$, which are not directly obtained from DDR on the NP M5-brane. 

\section{Symmetry}
\label{3}
\subsection{Gauge Symmetry}
The gauge transformation of fields is
\bea
\delta_{\Lambda}\Phi
&=&g \kappa^{\dot\rho}\partial_{\dot\rho}\Phi
\qquad
(\Phi = X^I, \Psi),\\
\delta_{\Lambda} b_{\a\dot{\sigma}}
&=&\partial_\a\Lambda_{\dot\sigma}
-\partial_{\dot\sigma}\Lambda_\a
+g\kappa^{\dot\tau}\partial_{\dot\tau}b_{\a\dot\sigma}
+g(\partial_{\dot\sigma}\kappa^{\dot\tau})b_{\a\dot\tau},\\
\delta_{\Lambda} b_{2\dot{\sigma}}
&=&-\partial_{\dot\sigma}\Lambda_{2}
+g\kappa^{\dot\tau}\partial_{\dot\tau}b_{2\dot\sigma}
+g(\partial_{\dot\sigma}\kappa^{\dot\tau})b_{2\dot\tau},\label{gt5}\\
\delta_{\Lambda} b^{\dm}&=&
\kappa^{\dm}+g\kappa^{\dn}\del_{\dn}b^{\dm}.\label{transf-bdm}
\eea
We expect that $U(1)$ gauge symmetry on D4-brane
can be examined from the gauge transformations on the NP M5-brane theory \cite{M51,M52}.
The gauge transformation parameter $\Lambda_2$
shall be identified with the $U(1)$ gauge transformation parameter.
This is consistent with the identification of $a_{\dm}$ with $b_{\dm 2}$.
The gauge symmetry parameterized by $\Lambda_{\dm}$, i.e.
volume-preserving diffeomorphism (VPD) symmetry still appears in the D4-brane.
Hence, we have gauge transformation on $a_{\dm}$,
\bea
\delta_{\Lambda} a_{\dm}&=&
\del_{\dm}\lambda+
g(\kappa^{\dn}\del_{\dn}a_{\dm}+a_{\dn}\del_{\dm}\kappa^{\dn}).\label{transf-adm}
\eea
The gauge symmetry combines $U(1)$ gauge symmetry and VPD symmetry.

\subsubsection{Gauge Transformation of $a_{\a}$}
The field $a_{\a}$ was introduced by the dual transformation and
its gauge transformation rule has to be solved from
the requirement that the dual action to be invariant.
First, we need to realize that Chern-Simons term
must be gauge invariant by itself.
The gauge transformation
\footnote{
The gauge transformation of $\breve{B}_{\a}^{~\dm}$
is the same as $B_{\a}^{~\dm}$. ($~\d_{\Lambda} \breve{B}_{\mu}{}^{\dm} =
\del_{\mu}\kappa^{\dm} + g\kappa^{\dn}\del_{\dn}\breve{B}_{\mu}{}^{\dm}
- g(\del_{\dn}\kappa^{\dm})\breve{B}_{\mu}{}^{\dn}$.)
}
 of the Chern-Simons term is
\bea
&&\delta_{\Lambda}\left(\eps^{\a\b}\del_{\b}a_{\dm}\breve{B}_{\a}^{~\dm}
+\frac{g}{2}\eps^{\a\b}F_{\dm\dn}\breve{B}_{\a}^{~\dm}\breve{B}_{\b}^{~\dn}
-\eps^{\a\b}\del_{\dm}a_{\b}\breve{B}_{\a}^{~\dm}\right)\nn\\
&=&\del_{\dm}\breve{B}_{\a}^{~\dm}\eps^{\a\b}\bigg[-\del_{\b}\lam
-g\left(\kappa^{\ds}\del_{\ds}a_{\b}+a_{\ds}\del_{\b}\kappa^{\ds}\right)
+ \d_{\Lambda} a_\b\bigg].
\eea
Hence we get
\be
\delta_{\Lambda}a_{\b}=\del_{\b}\lam
+g\left(\kappa^{\ds}\del_{\ds}a_{\b}+a_{\ds}\del_{\b}\kappa^{\ds}\right).
\label{transf-aa}
\ee

In our formulation of the self dual gauge fields $b$,
the components $b_{\mu\nu}$ do not explicitly appear in the action.
In \cite{Pasti,Furuuchi:2010sp},
the components $b_{\mu\nu}$ are used to explicitly exhibit
the self duality of the gauge field,
and their gauge transformation are given by
\be
\delta_{\Lambda}b_{\m\n}=\del_{\m}\Lambda_{\n}-\del_{\n}\Lambda_{\m}
+g\bigg[\kappa^{\dr}(\del_{\dr}b_{\m\n})
+(\del_{\n}\kappa^{\dr})b_{\m\dr}-(\del_{\m}\kappa^{\dr})b_{\n\dr}\bigg].
\ee
Identifying $b_{\b 2}$ with $a_{\b}$ and setting $\del_2 = 0$ from DDR,
we get exactly the same gauge transformation rule as eq.(\ref{transf-aa})
with $\Lambda_2 = \lam$.

We find the gauge transformation of $a_{\dm}$ (eq.(\ref{transf-adm}))
and $a_{\a}$ (eq.(\ref{transf-aa})) are of the same form ($A=\a,\dm$)
\be
\delta_{\Lambda} a_{A}=
\del_{A}\lambda+g\left(\kappa^{\dn}\del_{\dn}a_{A}+a_{\dn}\del_{A}\kappa^{\dn}\right).\label{transf-a}
\ee

Let us also give
the gauge transformation of $V_{\dn}{}^{\dm}$,
$\textbf{M}_{\dm\dn}{}^{\a\b}$, $W_{\dm}{}^{\a}$ and $\hat{B}_{\a}^{~\dm}$,
\bea
\delta_{\Lambda} V_{\dn}{}^{\dm} &=&
g\kappa^{\dlam}\del_{\dlam}V_{\dn}{}^{\dm}
+ g(\del_{\dn}\kappa^{\dlam}) V_{\dlam}{}^{\dm},
\\
\delta_{\Lambda} \textbf{M}_{\dm\dn}{}^{\a\b} &=&
g\bigg[\kappa^{\ds}\del_{\ds}\textbf{M}_{\dm\dn}{}^{\a\b}
+(\del_{\dm}\kappa^{\ds})\textbf{M}_{\ds\dn}{}^{\a\b}
+(\del_{\dn}\kappa^{\ds})\textbf{M}_{\dm\ds}{}^{\a\b}\bigg],
\\
\delta_{\Lambda} W_{\dm}{}^{\a}&=&\del_{\b}\kappa^{\ds}\textbf{M}_{\dm\ds}{}^{\a\b}+
g\bigg[\kappa^{\ds}\del_{\ds}W_{\dm}{}^{\a}+\del_{\dm}\kappa^{\ds}W_{\ds}{}^{\a}\bigg],
\\
\delta_{\Lambda} \hat{B}_{\a}^{~\dm} &=&
\del_{\a}\kappa^{\dm}+
g\left(\kappa^{\dn}\del_{\dn}\hat{B}_{\a}^{~\dm}-
\hat{B}_{\a}^{~\dn}\del_{\dn}\kappa^{\dm}\right).
\eea

\subsubsection{Covariant Variable with $U(1)$ and VPD Symmetry}
In the original NP M5-brane theory,
we have the covariant field strengths
\footnote{
A field $\hat{\Phi}$ is covariant if its gauge transformation is
$\delta_{\Lambda} \hat{\Phi} = g\kappa^{\dm}\del_{\dm}\hat{\Phi}$.
}
\bea
{\cal H}_{\dot{1}\dot{2}\dot{3}}&=&
\del_{\dm}b^{\dm}+\frac{1}{2}g
(\del_{\dn}b^{\dn}\del_{\dr}b^{\dr}-\del_{\dn}b^{\dr}\del_{\dr}b^{\dn})
+g^2 \{b^{\dot{1}},b^{\dot{2}},b^{\dot{3}}\}, \\
{\cal F}_{\dm\dn}&\equiv& {\cal H}_{2\dm\dn} =
F_{\dm\dn}+g
\bigg[\del_{\ds}b^{\ds}F_{\dm\dn}-\del_{\dm}b^{\ds}F_{\ds \dn}
-\del_{\dn}b^{\ds}F_{\dm\ds}\bigg].
\eea

The covariant version of $F_{\a\dm}$ can be defined as
\be
{\cal F}_{\a\dm} \equiv
\frac{1}{2}\eps_{\b\a}\eps_{\dm\dn\dlam}{\cal H}^{\b\dn\dlam}.
\ee
This is motivated by the intuition that ${\cal F}_{\a\dm}$
corresponds to ${\cal H}_{\a\dm 2}$ in the NP M5-brane theory.
Replacing $B_{\a}{}^{\dm}$ by $\hat{B}_{\a}{}^{\dm}$,
we can rewrite ${\cal H}^{\b\dn\dlam}$ from Eq.(\ref{h12def}) by the other fields.
(That is, we avoided to use $\del_{\a} b^{\dm}$ directly.
The dependence on $\del_{\a} b^{\dm}$ only appears through $\hat{B}_{\a}{}^{\dm}$.)
As a result, we have
\be
{\cal F}_{\a\dm}\equiv
{V^{-1}}_{\dm}^{~\dn}\left\{F_{\a\dn}+
g\bigg[F_{\dn\dd}\hat{B}_{\a}^{~\dd}+\eps_{\a\b}\del_{\dn}X^{I}\del^{\b}X^{I}-
\frac{i}{2}\eps_{\a\b}\bar{\Psi}\G^{\b}\del_{\dn}\Psi\bigg]
-g^{2}\eps_{\a\b}\hat{B}^{\b}_{\dr}\del_{\dn}X^{I}\del^{\dr}X^{I}\right\}.
\ee
This is also in agreement with the definition of ${\cal H}_{\mu\nu\dm}$
defined in \cite{Pasti,Furuuchi:2010sp}.

By inspection,
we can guess the covariant form of $F_{\a\b}$.
\bea
{\cal F}_{\a\b}&=&
F_{\a\b}+g\bigg[-F_{\a\dm}\hat{B}_{\b}^{~\dm}-
F_{\dm\b}\hat{B}_{\a}^{~\dm}+
gF_{\dm\dn}\hat{B}_{\a}^{~\dm}\hat{B}_{\b}^{~\dn}\bigg],\label{Fab}
\eea
where
\be
F_{AB} \equiv \del_A a_B - \del_B a_A.
\ee
Unlike ${\cal F}_{\dm\dn}$ and ${\cal F}_{\a\dm}$,
the components ${\cal F}_{\a\b}$ cannot be directly
matched with the field strength ${\cal H}_{\a\b 2}$ on M5-brane theory
because it involves the fields that
does not exist in D4-brane.

\subsection{Supersymmetry Transformation}

The supersymmetry transformation of fields (except for $a_{\a}$ and $\breve{B}_{\b}^{~\dn}$) on D4-brane from DDR is
\begin{eqnarray}
\label{eq:3.1}
\delta_{\eps} X^I &=&i\ol\epsilon\Gamma^I\Psi,\\
\delta_{\eps} \Psi
&=&{\cal D}_\a X^I\Gamma^\a\Gamma^I\epsilon
+{\cal D}_{\dot\mu}X^I\Gamma^{\dot\mu}\Gamma^I\epsilon+\frac{1}{2}g\eps^{\dm\dn\dr}F_{\dn\dr}\del_{\dm}X^{I}\G^{2}\G^{I}\epsilon
\nonumber\\&&
-\frac{1}{2}{\cal F}_{\dn\dr}\G^{2}\G^{\dn\dr}\epsilon-\frac{1}{2}
{\cal H}_{\a\dot\nu\dot\rho}
\Gamma^\a\Gamma^{\dot\nu\dot\rho}\epsilon
-\left(\frac{1}{g}+{\cal H}_{\dot1\dot2\dot3}\right)
\Gamma_{\dot1\dot2\dot3}\epsilon
\nonumber \\&&
-\frac{g^2}{2}\{X^{\dot\mu},X^I,X^J\}
\Gamma_{\dot\mu}\Gamma^{IJ}\epsilon
+\frac{g^2}{6}\{X^I,X^J,X^K\}
\Gamma^{IJK}\Gamma^{\dot1\dot2\dot3}\epsilon,\\
\delta_{\eps} b_{\dot\mu\dot\nu}
&=&-i(\ol\epsilon\Gamma_{\dot\mu\dot\nu}\Psi),\\
\delta_{\eps} a_{\dot\mu}
&=&i(\ol\epsilon\Gamma_2\Gamma_{\dot\nu}\Psi)(\delta_{\dm}{}^{\dn}+g\del_{\dm}b^{\dn})
-ig(\ol\epsilon\Gamma_2\Gamma^I\Gamma_{\dot1\dot2\dot3}\Psi)
\partial_{\dot\mu}X^I.
\end{eqnarray}
\subsubsection{Supersymmetry Transformation of ${\a_{\alpha}}$ and $\breve{B}_{\beta}^{~\dm}$}
We cannot obtain the supersymmetry transformation for ${\a_{\alpha}}$ and $\breve{B}_{\beta}^{~\dm}$ directly from DDR. But we find -$\eps^{\a\b}\del_{\dm}a_{\b}\breve{B}_{\a}^{~\dm}$ this term in action (Eq.(\ref{eq:2.1})) can give us some information to determine the supersymmetry transformation for $\breve{B}_{\beta}^{~\dm}$.
Firstly, the supersymmetry variation of $-\eps^{\a\b}\del_{\dm}a_{\b}\breve{B}_{\a}^{~\dm}$ will become
\begin{eqnarray}
\label{eq:3.2}
-\eps^{\a\b}\del_{\dm}\delta_{\eps} a_{\b}\breve{B}_{\a}^{~\dm}-\eps^{\a\b}\del_{\dm}a_{\b}\delta_{\eps}\breve{B}_{\a}^{~\dm}.
\end{eqnarray}
Because the other terms in action do not create $a_{\a}$ to cancel the second term of Eq.(\ref{eq:3.2}) from supersymmetry transformation, the only one way is to use partial integration by part to delete it (so it needs $\del_{\dm}\delta_{\eps}\breve{B}_{\a}^{~\dm}=0$). Fortunately, we have one simple candidate for
$\delta_{\epsilon}\breve{B}_{\mu}^{~\dm}$ is
$\epsilon^{\dot{\mu}\dot{\nu}\dot{\rho}}\partial_{\dot{\nu}}\delta_{\epsilon}b_{\mu\dot{\lambda}}$ (We already know supersymmetry transformation for $b_{\mu\dot{\lambda}}$ on M5-brane.).
From the above discussion, we know
\begin{eqnarray}
\label{eq:3.3}
\delta_{\eps}\breve{B}_{\a}^{~\dm}=-i\eps^{\dm\dn\dr}\ol\epsilon\G_{\a}\G_{\dlam}\del_{\dn}\Psi(\d_{\dr}{}^{\dlam}+g\del_{\dr}b^{\dlam})+ig\eps^{\dm\dn\dr}\ol\epsilon\G_{\a}\G^{I}\Gamma_{\dot1\dot2\dot3}\del_{\dn}\Psi
\partial_{\dr}X^I.
\end{eqnarray}
After many trivial but long calculations, we calculate
\begin{eqnarray}
\label{eq:3.4} \delta_{\epsilon}S&=&-\frac{1}{2}ig\delta_{\eps}\ol\Psi\G^{\a}\Psi\del_{\dm}\breve{B}_{\a}^{~\dm}-ig\eps^{\a\g}\ol\epsilon\Gamma_2\Gamma_{\dot\nu}\Psi\del_{\g}b^{\dn}\del_{\dm}\breve{B}_{\a}^{~\dm}\nn\\
&&+ig\eps^{\a\g}\ol\epsilon\Gamma_2\Gamma^I\Gamma_{\dot1\dot2\dot3}\Psi
\partial_{\g}X^I\del_{\dm}\breve{B}_{\a}^{~\dm}\nn\\
&&+\eps^{\a\g}\delta_{\eps} a_{\g}\del_{\dm}\breve{B}_{\a}^{~\dm}.
\end{eqnarray}
Hence, we can get supersymmetry transformation of $a_{\beta}$,
\begin{eqnarray}
\delta_{\eps} a_{\b}&=&-\frac{1}{2}ig\delta_{\eps}\ol\Psi\G^{\a}\Psi\eps_{\a\b}+ig\ol\epsilon\Gamma_2\Gamma_{\dot\nu}\Psi\del_{\b}b^{\dn}\nn\\
&&-ig\ol\epsilon\Gamma_2\Gamma^I\Gamma_{\dot1\dot2\dot3}\Psi\partial_{\b}X^I,
\end{eqnarray}
where
\begin{eqnarray}
\delta_{\eps} \ol\Psi
&=&\ol\epsilon\Gamma^I\Gamma^A{\cal D}_A X^I
+\frac{1}{2}g\ol\epsilon\G^{I}\G^{2}\eps^{\dm\dn\dr}F_{\dn\dr}\del_{\dm}X^{I}
\nonumber\\&&
-\frac{1}{2}\ol\epsilon\G^{\dn\dr}\G^{2}{\cal F}_{\dn\dr}-\frac{1}{2}\ol\epsilon\Gamma^{\dot\nu\dot\rho}\Gamma^\a
{\cal H}_{\a\dot\nu\dot\rho}
-\ol\epsilon\Gamma_{\dot1\dot2\dot3}\left(\frac{1}{g}+{\cal H}_{\dot1\dot2\dot3}\right)
\nonumber\\&&
-\frac{g^2}{2}\ol\epsilon\Gamma^{IJ}\Gamma_{\dot\mu}\{X^{\dot\mu},X^I,X^J\}
+\frac{g^2}{6}\ol\epsilon\Gamma^{\dot1\dot2\dot3}\Gamma^{IJK}\{X^I,X^J,X^K\}.
\end{eqnarray}
\subsubsection{Linearized Supersymmetry Transformation}
This theory has $16$ additional fermionic symmetries
$\delta_{\chi}$,
which shifts the fermion by a constant spinor
\begin{equation}
\delta_{\chi}\Psi=\chi,\quad
\delta_{\chi}X^I=\delta_{\chi}b^{\dot\mu}
=\delta_{\chi}a_{\dm}=0.
\end{equation}
We also get this fermionic transformation of $a_{\alpha}$,
\begin{equation}
\label{eq:3.6}
\delta_{\chi}a_{\a}=-\frac{i}{2}g\ol\chi\G^{\b}\Psi\eps_{\a\b}.
\end{equation}
These two supersymmetry transformations ($\d_{\eps}$, $\d_{\chi}$) are all nonlinear, which means that the supersymmetry transformations have constant spinor terms. When all fields vanish, the transformation does not vanish. This kind of SUSY is actually a broken SUSY. This result comes from the background effect.
After excluding the background effect, we can linearize the supersymmetry transformation,
\begin{equation}
\label{eq:3.7}
\delta\equiv\frac{1}{g}\delta_{\chi\rightarrow\Gamma_{\dot1\dot2\dot3}\eps}+\delta_{\eps},
\end{equation}
then we get linearized supersymmetry transformation of $a_{\a}$,
\begin{equation}
\label{eq:3.8}
\delta a_{\a}=\delta a_{\dm\rightarrow\a}+\frac{1}{2}\delta\ol\Psi\G^{\b}\Psi\eps_{\a\b}.
\end{equation}
On the other hand, the $\d\Psi$ do not have $\frac{1}{g}$ term,
\bea
\delta \Psi
&=&{\cal D}_\a X^I\Gamma^\a\Gamma^I\epsilon
+{\cal D}_{\dot\mu}X^I\Gamma^{\dot\mu}\Gamma^I\epsilon+\frac{1}{2}g\eps^{\dm\dn\dr}F_{\dn\dr}\del_{\dm}X^{I}\G^{2}\G^{I}\epsilon
\nonumber\\&&
-\frac{1}{2}{\cal F}_{\dn\dr}\G^{2}\G^{\dn\dr}\epsilon-\frac{1}{2}
{\cal H}_{\a\dot\nu\dot\rho}
\Gamma^\a\Gamma^{\dot\nu\dot\rho}\epsilon
-{\cal H}_{\dot1\dot2\dot3}\Gamma_{\dot1\dot2\dot3}\epsilon
\nonumber \\&&
-\frac{g^2}{2}\{X^{\dot\mu},X^I,X^J\}
\Gamma_{\dot\mu}\Gamma^{IJ}\epsilon
+\frac{g^2}{6}\{X^I,X^J,X^K\}
\Gamma^{IJK}\Gamma^{\dot1\dot2\dot3}\epsilon. \label{dPsi}
\eea

\subsubsection{Supersymmetry Transformation of $\hat{B}_{\a}{}^{\dm}$ Field}

Since Eq.(\ref{eq:2.1}) is at most quadratic in $\breve{B}_{\a}{}^{\dm}$ so we can do Gaussian
integration. Classically, it is equivalent to replace the field $\breve{B}_{\a}{}^{\dm}$ in action by its solution of EOM ($\hat{B}_{\a}{}^{\dm}$). Now we worry the supersymmetry transformation of $\hat{B}_{\a}{}^{\dm}$ may be different. In fact, $\d_{\eps}\hat{B}_{\a}{}^{\dm}$ is equal to $\d_{\eps}\breve{B}_{\a}{}^{\dm}$ with the additional terms proportional to EOM of
some fields. The difference what we get is,
\bea
\d_{\eps}\hat{B}_{\a}{}^{\dm} &=& \d_{\eps}\breve{B}_{\a}{}^{\dm} - 2 (\textbf{M}^{-1})^{\dm\dn}_{~~~\a\b}(\d_{\eps}\ol\Psi|_{\hat{B}})_{\dn}{}^{\b}(\mbox{EOM of}~\ol\Psi),\label{dhatB}
\eea
where
\be
(\d_{\eps}\ol\Psi|_{\hat{B}})_{\dn}{}^{\b} \equiv \frac{1}{2}\ol\eps\G^{\dr\dlam}\G^{\b}\eps_{\ds\dr\dlam}V_{\dn}{}^{\ds}-g\ol\eps\G^{I}\G^{\b}\del_{\dn}X^{I},
\ee
which is the coefficient of $\hat{B}_{\dn}{}^{\b}$ in $\d_{\eps}\ol\Psi$. \\The EOM of $\ol\Psi$ is
\bea
(\mbox{EOM of}~\ol\Psi)&=&\frac{i}{2}\G^{\a}\del_{\a}\Psi+\frac{i}{4}g\G^{2}\eps^{\dm\dn\dr}F_{\dn\dr}\del_{\dm}\Psi-\frac{i}{2}g\G^{\a}\hat{B}_{\a}{}^{\dm}\del_{\dm}\Psi\nn\\
&&+\frac{i}{2}\G^{\dr}{\cal D}_{\dr}\Psi+\frac{i}{2}g^{2}\G_{\dm}\G^{I}\{X^{\dm},X^{I},\Psi\}-\frac{i}{4}g^2\G^{IJ}\G^{\dot{1}\dot{2}\dot{3}}\{X^{I},X^{J},\Psi\}.\nn\\
\eea
This relation (\ref{dhatB}) also implies
\be
\d_{\eps}(\textbf{M}_{\dm\dn}^{~~\a\b}\hat{B}_{\b}{}^{\dn}-W_{\dm}^{\a})=0,
\ee
which means that we can write down a simpler form than Eq.(\ref{dhatB}). This simpler form can be understood as the $\d_{\eps}\breve{B}_{\a}{}^{\dm}$ with the additional terms proportional to EOM of fermionic field.
On the other hand, the another fermionic symmetry of $\hat{B}_{\a}{}^{\dm}$ is easy to calculate,
\be
\d_{\chi}\hat{B}_{\a}{}^{\dm}=0.
\ee

After integrating out $\breve{B}_{\a}{}^{\dm}$ field, the supersymmetry transformation of fermion is just to replace $\breve{B}_{\a}{}^{\dm}$ with $\hat{B}_{\a}{}^{\dm}$. Now we get all supersymmetry transformation of fields on D4-brane in the large $C$-field background. We hope to use the handle to open the new directions for the researches of D$p$-brane in the large $(p-1)$-form background in the future.

\subsection{Super Algebra and Central Charges}
To compute super algebra, we start to calculate the super charge of the D4-brane in the large $C$-field background
\footnote{In this section, we follow the methods mentioned in the papers \cite{Lambert,Low}.} \footnote{In the papers \cite{Low,Low2}, the author calculates the central charges from the superalgebra of the NP M5-brane theory, and performs DDR to get the central charges of D4-brane theory in the large $B$-field background.  He also provides more details with the interpretation of these central charges.}. The super charge $Q$ is calculated from the spatial integral of the time component of supercurrent $J^{0}$, where $J^{0}$ is defined by this way $\ol{\eps}J^{0}=-\d\ol{\Psi}\G^{0}\Psi$. 
The anticommutator of the supercharges is
\bea
\{Q,Q\}=2\int d^{4}x\; T_{00}+\sum_{n=0}^{5}\int d^{4}x\;Z_{n}.
\eea
Here, we divide it into two parts: the energy part and the central charges part.

The energy part is
\bea
T_{00}&=&\frac{1}{2}{\cal D}_{0}X^I{\cal D}_{0}X^I
+\frac{1}{2}({\cal D}_{a}X^I)^2+\frac{1}{4}g^{2} F_{\dm\dn}F^{\dm\dn}\del_{\dr}X^{I}\del^{\dr}X^{I}+\frac{1}{2}g^{2}F_{\dm\dn}F^{\dn\dr}\del_{\dr}X^{I}\del^{\dm}X^{I} \nn \\ 
&&
 +\frac{1}{4}{\cal H}_{0\dot\mu\dot\nu}{\cal H}_{0}{}^{\dot\mu\dot\nu}+\frac{1}{4}({\cal H}_{1\dot\mu\dot\nu})^2
+ \frac{1}{4}({\cal F}_{\dot\mu\dot\nu})^2
+\frac{1}{2}({\cal H}_{\dot1\dot2\dot3})^2  \nn \\ 
&&
+\frac{g^4}{4}\{X^{\dot\mu}, X^I, X^J\}^2
+\frac{g^4}{12}\{X^I, X^J, X^K\}^2.
\eea

We classify the central charges part according to the number of the scalar fields we choose to turn on.
In order to compare with the BPS solutions in the next section, the momentum term ($T_{0a}, a=(1,\dot\mu)$) is included in the central charges.
Let us now describe each terms of $Z_n$.
The convention of indices here are $\bar{a}=(0, \dot\mu)$.
\bea
Z_{0}={\cal H}_{0\dot\mu\dot\nu}{\cal H}^{\dot\mu\dot\nu a}
\Gamma^0\Gamma_{a}
-{\cal H}_{0\dot\nu\dot\rho}{\cal F}^{\dot\nu\dot\rho}\Gamma^2\Gamma^0
+{\cal F}_{\dot\rho\dot\lambda}{\cal H}_{1\dot\sigma\dot\omega}\Gamma^{\dot\rho}\Gamma^{\dot\lambda\dot\sigma\dot\omega}\Gamma^{12}.
\eea
We find ${\cal H}$${\cal H}$ on M5-brane becomes ${\cal F}$${\tilde{\cal F}}$ and ${\cal H}$${\cal H}$ on D4-brane.
From the D4-brane perspective,
${\cal F}$${\tilde{\cal F}}$ can be thought as the charge of a D0-brane
within the worldvolume of the D4-brane and
${\cal H}$${\cal H}$ can be thought as a pp-wave intersected a D4-brane.

Next, we have
\bea
Z_{1}&=&-g\epsilon^{\dot\mu\dot\nu\dot\rho}F_{\dot\nu\dot\rho}\partial_{\dot\mu}X^ID_0X^I\Gamma^2\Gamma^0+
2D_0X^ID_{a}X^I\Gamma^0\Gamma^{a}+\frac{1}{3}D_{a}X^I{\cal H}_{bcd}\Gamma^{abcd}\Gamma^I\nn\\
&&+\frac{g}{6}\epsilon^{\dot\mu\dot\nu\dot\rho}F_{\dot\nu\dot\rho}\partial_{\dot\mu}X^I{\cal H}_{bcd}\Gamma^{2bcd}\Gamma^I
+D_{a}X^I{\cal F}_{\dot\nu\dot\rho}\Gamma^{a 2\dot\nu\dot\rho}\Gamma^I\nn\\
&&+D_{\dot\mu}X^I{\cal H}_{0\dot\lambda\dot\sigma}\epsilon^{\dot\nu\dot\lambda\dot\sigma}\Gamma^{21\dot\mu\dot\nu}\Gamma^I.
\eea
The $({\cal D}X){\cal H}$ term corresponds to the charge of F1 ending on D4 solution.
If we consider
one scalar field is active and assume that this scalar is a
function of only four of the spatial worldvolume coordinates of the D4-brane.
On the other hand, $\frac{g}{6}\epsilon^{\dot\mu\dot\nu\dot\rho}F_{\dot\nu\dot\rho}\partial_{\dot\mu}X^I{\cal H}_{bcd}$ and $D_{\dot\mu}X^I{\cal F}_{\dot\nu\dot\rho}$ terms correspond to our BPS solution with two types of magnetic charges.

When two scalar fields are turned on, we will need to consider $Z_{0}$, $Z_{1}$ and $Z_{2}$, where $Z_{2}$ is defined by
\bea
Z_{2}&=&D_aX^ID_bX^J\Gamma^{ab}\Gamma^{IJ}
+g\epsilon^{\dot\mu\dot\nu\dot\rho}F_{\dot\nu\dot\rho}\partial_{\dot\mu}X^ID_bX^J\Gamma^{2b}\Gamma^{IJ}
+2g^2D_{\dot\mu}X^I\{X^{\dot\mu},X^I,X^J\}\Gamma^J\nn\\
&&+2g^2D_0X^I\{X^{\dot\nu},X^I,X^J\}\Gamma^0\Gamma_{\dot\nu}\Gamma^J
+\frac{g^2}{2}{\cal H}_{0\dot\nu\dot\lambda}\{X_{\dot\mu},X^J,X^K\}\Gamma^{\dot\mu\dot\nu\dot\lambda}\Gamma^{JK}\Gamma^0\nn\\
&&-g^2{\cal H}_{1\dot\rho\dot\nu}\{X^{\dot\nu},X^I,X^J\}\Gamma^{1\dot\rho}\Gamma^{IJ}
-g^2{\cal F}_{\dot\rho\dot\nu}\{X^{\dot\nu},X^I,X^J\}\Gamma^{2\dot\rho}\Gamma^{IJ}\nn\\
&&-g^2{\cal H}_{\dot1\dot2\dot3}\{X^{\dot\mu},X^I,X^J\}\Gamma_{\dot\mu}\Gamma^{IJ}\Gamma^{\dot1\dot2\dot3}.
\eea
We see the first term ${\cal D}_a X^I{\cal D}_b X^J$ corresponds
to the charge of the 2-brane vortex living on the D4-brane worldvolume.

We also have
\bea
Z_{3}&=&g^2D_{a}X^{I}\{X^{\dot\mu},X^J,X^K\}\Gamma^{a}\G_{\dot\mu}\Gamma^{IJK}
+\frac{g^3}{2}\epsilon^{\dot\mu\dot\nu\dot\rho}
F_{\dot\nu\dot\rho}\partial_{\dot\mu}X^I\{X^{\dot\lambda},X^J,X^K\}\Gamma^{2}\G_{\dot\lambda}\Gamma^{IJK}\nn\\
&&-g^2D_{\bar{a}}X^I\{X^I,X^J,X^K\}\Gamma^{\bar{a}}\Gamma^{JK}\Gamma^{\dot1\dot2\dot3}
-\frac{g^2}{6}{\cal H}_{1\dot\rho\dot\lambda}\{X^I,X^J,X^K\}\Gamma^{1\dot\rho\dot\lambda}
\Gamma^{IJK}\Gamma^{\dot1\dot2\dot3}\nn\\
&&-\frac{g^2}{6}{\cal F}_{\dot\rho\dot\lambda}\{X^I,X^J,X^K\}\Gamma^{2\dot\rho\dot\lambda}\Gamma^{IJK}\Gamma_{\dot1\dot2\dot3}
+g^4\{X^I,X^J,X^{\dot\mu}\}\{X^I,X^K,X^{\dot\nu}\}\Gamma_{\dot\mu\dot\nu}\Gamma^{JK}\nn\\
\eea
and
\bea
Z_4&=&-\frac{g^2}{3}D_1X^I\{X^J,X^K,X^L\}\Gamma^1\Gamma^{IJKL}\Gamma^{\dot1\dot2\dot3}\nn\\
&&-\frac{g^3}{6}\epsilon^{\dot\mu\dot\nu\dot\rho}F_{\dot\nu\dot\rho}\partial_{\dot\mu}X^I\{X^J,X^K,X^L\}\Gamma^2
\Gamma^{IJKL}\Gamma_{\dot1\dot2\dot3}\nn\\
&&+g^4\{X^{\dot\mu},X^I,X^J\}\{X^I,X^K,X^L\}\Gamma_{\dot\mu}\Gamma^{JKL}\Gamma_{\dot1\dot2\dot3}\nn\\
&&-\frac{g^4}{4}\{X^{\dot\mu},X^I,X^J\}\{X_{\dot\mu},X^K,X^L\}\Gamma^{IJKL}.
\eea
The charge $\frac{g^2}{3}{\cal D}_1 X^I\{X^J, X^K, X^L\}$
is equipped with 3-bracket on D4-brane.
The geometry of the M5-brane is similar with the situation of the $C$-field modified Basu-Harvey equation as a boundary condition of the multiple M2-brane theory \cite{Chu:2009iv}.

Finally, the last of $Z_{n}$ is
\bea
Z_{5}=-\frac{g^4}{4}\{X^I, X^J, X^K\}\{X^I, X^L, X^M\}\Gamma^{JKLM}.
\eea
This term is relevant only if we turn on all scalars $X^{6},\cdots, X^{10}$. 

\section{BPS Solutions}
\label{4}
In this paper, we only consider pure bosonic solitons,
namely those with the fermion field $\Psi = 0$.
The BPS condition is therefore simply that
the supersymmetry transformation of $\Psi$ (eq.(\ref{dPsi})) vanishes
for some supersymmetry parameters $\eps$.
We systematically study BPS solutions
by classifying them according to the number of scalars
that are turned on. 

Most of these BPS solutions are directly derived from the NP M5-brane by DDR. So we do not write down the details of these solutions.
We divide two parts of this section here. In the first part, we mention the BPS solutions on D4-brane which are derived from the NP M5-brane directly and discuss the instanton solution particularly. The next part is to discuss the BPS solutions after we turn on one scalar field ($X^{6}$). This solution is called BPS solution with two types of magnetic charges, which can be understood as D2 ending on D4 in geometric viewpoint. This solution can be related to self-dual string solution on M5-brane from \lq\lq{}non-linear superpositions\rq\rq{}. In order to emphasize it, we put this solution in a new section.

\subsection{Solutions via DDR from the NP M5-brane BPS States}

Some solutions are easy to be derived from the NP M5-brane by DDR, so we do not write down the details from these solutions. From the NP M5-brane theory, we have obtained M-waves, the self-dual string (M2 ending on M5), tilted M5-brane, holomorphic embedding of M5-brane and the intersection of two M5-branes along a 3-brane BPS solutions. We classify these solutions in the previous paper \cite{Ho:2012dn} with the number of scalar fields we turn on. When we turn off all scalar fields, we get M-waves (Light-Like) BPS solutions. If we turn on one scalar field $X^{6}$, we get the self-dual string (M2 ending on M5) and tilted M5-brane BPS solutions. When we turn on two scalar fields $X^6$ and $X^7$, we get holomorphic embedding of M5-brane and the intersection of two M5-branes along a 3-brane BPS solutions.  

After doing double dimensional reduction, it directly obtains these BPS solutions on D4-brane from the above solutions. We can obtain pp-wave, F1 ending on D4 solution, tilted D4-brane, holomorphic embedding of D4-brane and the intersection of two D4-branes along a 2-brane.  The light-like BPS solutions in M5-brane theory give the pp-wave BPS solutions and the trivial instanton solution on D4-brane \footnote{The reason for this instanton solution is trivial will be given latter.}. The BPS solutions with turning on one scalar field $X^6$ on M5-brane can be related  directly to the F1 ending on D4 and the tilted D4-brane BPS solutions in D4-brane theory. Finally, the BPS solutions with two scalar fields on M5-brane can be reduced to the holomorphic embedding of D4-brane and the intersection of two D4-branes along a 2-brane BPS solutions in D4-brane theory.

\subsubsection{Instanton Solution}
Even if these solutions can easily be derived from BPS solutions in the NP M5-brane theory. There is still an interesting issue about the BPS stares of D0-brane (instanton) \footnote{We call ``instanton'' because we use the same terminology of D(-1) solutions in D3-brane case.} exists whether or not.  
In these solutions, we do not have a non-trivial instanton solution in the large $C$ field background after we do DDR.
To emphasize the reason, we show the detail calculation on this solution.  
The instanton solution on D4-brane can be related to the light-like BPS solution in the previous paper \cite{Ho:2012dn}. The BPS solution satisfies the BPS condition $\G^{02} \eps = \pm \eps$. After DDR, the BPS conditions are
\bea
&{\cal H}_{0\dm\dn} = \pm {\cal F}_{\dm\dn},
\\
&{\cal H}_{1\dm\dn} = 0,
\\
&{\cal H}_{\dot{1}\dot{2}\dot{3}} = 0.
\eea
This solution preserves $\frac{1}{2}$ SUSY, which can be thought as a D0-brane.
The energy density is bound by $\mid\frac{1}{2}{\cal H}_{0\dot\mu\dot\nu}{\cal F}^{\dot\mu\dot\nu}\mid$,
and it is consistent with the central charge $Z_{0}$.

Firstly, we impose a gauge fixing condition
\bea
b^{\dot1}=b^{\dot2}=b^{\dot3}=0.
\eea
From ${\cal H}_{1\dot\mu\dot\nu}$=0, we obtain
\bea
\hat{B}_{1}{}^{\dot\mu}=0.
\eea
From ${\cal H}_{0\dot\mu\dot\nu}=\pm{\cal F}_{\dot\mu\dot\nu}$, we obtain
\bea
\hat{B}_{0}{}^{\dot\mu}=\mp \frac{1}{2}\epsilon^{\dot\mu\dot\nu\dot\rho} F_{\dot\nu\dot\rho}.
\eea

Let us consider the equation of motion eq.(\ref{r2}) without turning on scalar,
\be
-\hat{B}^{\alpha}{}_{\dot\mu}+\epsilon^{\alpha\beta}F_{\beta\dot\mu}+g\epsilon^{\alpha\beta}F_{\dot\mu\dot\nu}
\hat{B}_{\beta}{}^{\dot\nu}=0.
\ee
This implies
\bea
\hat{B}_{0\dot\mu}=-F_{1\dot\mu}.
\eea
We can get the following relation,
\bea
&&F_{1\dot 1}=\pm F_{\dot2\dot3},\nn \\
&&F_{1\dot 2}=\pm F_{\dot3\dot1},\nn \\
&&F_{1\dot 3}=\pm F_{\dot2\dot1}.
\eea
As $U(1)$ gauge theory, we do not have non-trivial instanton solutions \footnote{In five dimensional theory, these relations can be understood as self-duality equations with static gauge fields and temporal gauge $a_{0}=0$.}. So we do not have non-trivial instanton solutions in the large $C$-field background \footnote{In fact, our effective theory is well defined in a special scaling limit \cite{Ho}, it is still possible that instanton solutions exist in other limits.}.

This BPS solution of M5-brane is a wave which travels at the speed of light on $x_2$ direction. If we do DDR on this direction, we will obtain the trivial solution on D4-brane in the large C-field background. However, we propose the next possible instanton solution from combining $\G^{02}\eps=\pm\eps$ and $\G^{026}=\pm\eps$. This solution preserves $\frac{1}{4}$ SUSY with D0-brane and F1-string ending on D4-brane interpretation.

\subsection{BPS Solution with Two Types of Magnetic Charges}
After turning on one scalar $X^{6}$, we have a new solution which is related to the solution of the NP M5-brane theory with the \lq\lq{}non-linear\rq\rq{} way. The BPS condition which we consider is $\Gamma^{016}\epsilon$=$\pm$$\epsilon$. This solution preserves $\frac{1}{2}$ SUSY, and the geometric picture is D2 ending on D4. The energy density is bounded by 
\be
\mid\frac{1}{2}\epsilon^{\dot\mu\dot\nu\dot\rho}\left({\cal F_{\dot\nu\dot\rho}}D_{\dot\mu}X^{6}-gF_{\dot\nu\dot\rho}{\cal H}_{\dot{1}\dot{2}\dot{3}}\partial_{\dot\mu}X^{6}\right)\mid,
\ee
and it is consistent with the central charges on D4-brane.

This solution up to first g order is \footnote{Here, $\breve{B}_{\a}{}^{\dm}=0$.},
\bea
X^{6}&=&\pm\bigg[\frac{m_{2}}{a}+g\frac{m_{1}m_{2}}{a^{4}}+{\cal O}(g^{2})\bigg],\\
b^{\dot\mu}&=&-\frac{m_{1}}{a^{3}}x^{\dot\mu}+g\left( -\frac{m_{1}^{2}}{a^{6}}+\frac{m_{2}^{2}}{a^{4}}\right)x^{\dot\mu}+{\cal O}(g^{2}), \label{p2}\\
F_{\dot\nu\dot\rho}&=&-\epsilon_{\dot\nu\dot\rho\dot\mu}\frac{m_{2}}{a^{3}}x^{\dot\mu}+{\cal O}(g^{2}),
\eea
where the notation $a$ is $\sqrt{x_{\dot1}^2 + x_{\dot2}^2 +x_{\dot3}^2 }$.

From this solution, we can know it contains two types of magnetic charges. The one ($Q_{M2}$=-4$\pi m_{2}$, where $m_{2}$=$\frac{k_{2}}{(2\pi)^{\frac{3}{2}}T_{D_{4}}^{\frac{2}{5}}}$) is from $F^{\dot\mu\dot\nu}$ and another ($Q_{M1}$=-4$\pi m_{1}$, where $m_{1}$=$\frac{k_{1}}{(2\pi)^{\frac{3}{2}}T_{D_{4}}^{\frac{3}{5}}}$) is from ${\cal H}_{\dot{1}\dot{2}\dot{3}}$. And we see two charges have the interaction from the results of the first-order expansion. The reason is due to ${\cal F}_{\dot\nu\dot\rho}$. This field strength offers interaction between $a^{\dot{\mu}}$ and $b^{\dot{\nu}}$ from the first-order term. This interaction is due to the strength of Nambu-Poisson bracket so this interaction disappears if $g$=0. We also find an interesting connection between this solution and solution of Ref. \cite{Furuuchi}, which is called Furruchi-Takimi (FT) solution in this paper. We find this solution is just a  \lq\lq{}non-linear superposition\rq\rq{} of FT solution, if we integrate FT solution with respect to $x_{2}$ \footnote{We do integration before we perform DDR on $x_2$ this direction so the range of $x_{2}$ is from $-\infty$ to $\infty$.},
\bea
\int X_{(0)}^{6(FT)}dx_{2}&=&\int\frac{m}{r^{2}}dx_{2}=\frac{m\pi}{a},\\
\int {\cal H}_{(0)\dot{1}\dot{2}\dot{3}}^{(FT)}dx_{2}&=&\int\frac{-2mx_{2}}{r^{4}}dx_{2}=0,
\eea
where $X_{(0)}^{6}{}^{(FT)}$ and ${\cal H}_{(0)}^{(FT)}$ are the zero-order of FT solutions.
When $m_{2}$=$m\pi$, two solutions are the same. We can interpret our zero-order solutions are just linear superposition of the zero-order of FT solutions. If we examine the first-order solutions, we find two solutions cannot be the same from integration. Due to the strength of Nambu-Poisson bracket breaks the linear superposition effect, we interpret this soliton solution is a \lq\lq{}non-linear superposition\rq\rq{} of FT solution.

We also find an ansatz from the perturbative solutions.
\bea
b^{\dot\mu}&=&f\left( a \right) x^{\dot{\mu}},\nn\\
X^{6}&=&h\left( a \right),\nn\\
F^{\dot\mu\dot\nu}&=&\epsilon^{\dot\mu\dot\nu\dot\rho}\frac{C}{a^{3}}x_{\dot\rho},
\eea
where $C$ is an arbitrary constant. From the perturbation, we observe $F^{\dot\mu\dot\nu}$ should be a function of $a$. And we have $dF=0$ this restriction. The only one possible ansatz should be monopole solution for field strength. It creates a source on the origin.
The function of $f(a)$ and $h(a)$ need to satisfy
\bea
\frac{df}{da}&=&-\frac{a^{4}f(3+3gf+g^{2}f^{2})(gf+1)^{2}+gC^{2}(gf+1)}{a\left( a^{4}(gf+1)^{4}+C^{2}g^{2}\right)},\\
\frac{dh}{da}&=&\frac{Ca^2}{a^{4}(1+gf)^{4}+C^{2}g^{2}}.
\eea
If we want $f$ and $h$ are smooth function, we need $gf(0)$+1=0 this constraint on the origin.
We find the only one smooth solution on this boundary condition, but it does not have finite energy.
\bea
b^{\dot\mu}&=&-\frac{1}{g}x^{\dot\mu},\nn\\
X^6&=&\frac{a^3}{3Cg^2}+\frac{E}{Cg^2},\nn\\
F^{\dot\mu\dot\nu}&=&\epsilon^{\dot\mu\dot\nu\dot\rho}\frac{C}{a^{3}}x_{\dot\rho},
\eea
where $E$ is an arbitrary constant. We think this ansatz is not good enough to probe the non-perturbative soliton solution and this solution should be the tilted D4-brane BPS solution. But it is still an interesting problem to find an exact solution to describe this soliton solution near origin in the future. 

\section{Conclusion and Discussion}
\label{5}

In this paper, we give more details of the low energy effective theory for D4-brane in the large $C$-field background. When theory couples with matter fields, there are not only U(1) and VPD gauge symmetry but also supersymmetry. We obtain the supersymmetry transformation of all fields in the low energy effective theory. In the previous paper \cite{Ho}, we did not point out how to calculate the supersymmetry transformation law after the duality transformation. After the duality transformation, the interpretation of $B_{\a}{}^{\dm}$ field was changed. We should treat the field $B_{\a}{}^{\dm}$ to be the new field $\breve{B}_{\a}{}^{\dm}$ in the dual action. So the divergence of $\breve{B}_{\a}{}^{\dm}$ does not vanish. ($\del_{\dm}\breve{B}_{\a}{}^{\dm} \neq 0$.) This property helps us to find the supersymmetry transformation of $a_{\a}$, which comes from the duality transformation. After obtaining all supersymmetry transformation laws, we also check the Lagrangian being supersymmetrical invariant. Now, the full supersymmetry of this effective action is completely understood in this work.

Moreover, we are also interested in the topological quantities of this theory so we calculated the central charges from supercurrent. These central charges let us know the possible topological solutions. In the last section, we studied BPS solutions of the effective field theory for D4-brane
in the large $C$-field background.
The large $C$-field background turns on new interactions
on the D4-brane worldvolume through
the Nambu-Poisson structure,
and modifies some of the BPS configurations. Most of them correspond to the double dimensional reduction of the BPS solutions in the NP M5-brane theory so we just mention them without details. On the other hand, we also found a new perturbative solution which was not directly related to the self-dual string BPS solution in the NP M5-brane theory \cite{Ho:2012dn, Furuuchi} after we do DDR. It is related to the self-dual string solution with \lq\lq{}non-linear superpositions\rq\rq{}. This geometric picture can be easily understood as D2 ending on D4.

We did not find the instanton solutions in this effective theory. Originally, we are interested in the topological quantities because of the well-known $U(1)$ instanton solution of D-brane in the large NS-NS $B$-field background \cite{Seiberg:1999vs, Nekrasov}. We wonder if there is a similar $U(1)$ instanton solution in the large $C$-field background. However, we cannot find it in $\frac{1}{2}$ BPS states. But this solution may survive in $\frac{1}{4}$ BPS solutions.

Finally, it is still an open question: how to generalize our work to all D$p$-brane in all R-R field backgrounds with matter fields. If the generalization is successful, it is possible to find new BPS states. These work help us to understand more about the geometrical structure of the Nambu-Poisson gauge theory and open the new direction on D$p$-brane in the R-R field background.

\section*{Acknowledgment}
The authors thank Prof. Pei-Ming Ho. This paper can be finished because of his encouragement and many useful suggestions.
The authors also thank Wei-Ming Chen, Kazuyuki Furuuchi, Hiroshi Isono, Sheng-Lan Ko and Tomohisa Takimi for useful discussions. Without their discussions, it is possible to lose some interesting ideas.
The authors are supported in part by the National Science Council, Taiwan, R.O.C.

\end{document}